\documentstyle[sprocl,epsfig,floatflt]{article}

\def\be{\begin{equation}}
\def\ee{\end{equation}}
\def\bea{\begin{eqnarray}}
\def\eea{\end{eqnarray}}

\begin{document}

\title{Spin dynamics in Cuprates and its relation to superconductivity}

\author{Ph. BOURGES}

\address{Laboratoire L\'eon Brillouin, CEA-CNRS, CE Saclay, 91191 Gif sur 
Yvette, France\\E-mail: bourges@bali.saclay.cea.fr} 

\maketitle\abstracts{The relevance of magnetism for 
the mechanism responsible for high-temperature superconductivity remains 
an open and still interesting issue. The observation by inelastic 
neutron scattering of strong antiferromagnetic dynamical correlations 
in superconducting cuprates is discussed in relation to the unusual physical 
properties of the cuprates as well as in relation to the superconducting 
pairing. }

\section{Why neutron scattering in High-$T_c$ Cuprates ?}

Although conventional electron-phonon interaction alone within 
BCS  \cite{bcs} theory can be dismissed,
fourteen years after its discovery \cite{muller}, the mechanism for 
high-temperature superconductivity in copper oxides is still debated. 
Among more exotic approach for superconductivity,  
spin fluctuations remain a serious candidate \cite{pines,afmodel}.
Indeed, the systematic existence of strong antiferromagnetic (AF) 
dynamical correlations in the metallic and superconducting phases
of all cuprates supports by principle that proposal.
The observation of magnetic fluctuations around the AF wavevector 
${\bf Q_{AF}} \equiv (\pi,\pi)$ was first emphasized by copper 
Nuclear Magnetic Resonance (NMR) \cite{nmr} and then widely reported 
by inelastic neutron scattering (INS). INS is actually playing an essential 
role on this matter as it is the only technique which 
directly measures the imaginary part of the spin susceptibility,
$Im \chi ({\bf Q},\hbar\omega)$, over a wide energy range 
($\hbar\omega \sim$ 1 to 200 meV) and for any momentum transfer 
within the Brillouin zone. [In contrast, spin lattice relaxation rate 
in NMR experiments only probes a sum of $Im \chi ({\bf Q},\hbar\omega)$ 
in momentum space weighting by atomic hyperfine tensor and at
frequencies $\omega \to 0$.] In principle, the full determination of the
spin susceptibility by neutron experiments would ultimately answer whether  
the  mechanism for high-temperature superconductivity is due to 
AF fluctuations or not. 

The powerfulness of inelastic neutron scattering is unfortunately limited 
by the need of large single crystals (of cm$^3$ size) usually difficult
to grow in complex systems such as high-$T_c$ cuprates. For that reason, 
only two cuprates families have been extensively studied by INS so far: 
${\rm La_{2-x}Sr_xCuO_{4}}$ (LSCO)~\cite{kastner,aeppli} and 
${\rm YBa_{2}Cu_{3}O_{6+x}}$ (YBCO) 
\cite{rossat91,tony,revue-cargese,revue-lpr,mook,dai99,tony2000,science,klosters}.
Further, although they have common features (energy scale, 
absolute units) \cite{revue-cargese}, the spin susceptibility observed in 
the two systems also exhibits clear differences: namely the 
low energy spin fluctuations are peaked in LSCO at wavevectors 
${\bf Q_{\delta}}=(\pi(1\pm\delta),\pi) \equiv (\pi,\pi(1\pm\delta))$, 
incommensurate from the AF momentum. In YBCO, the spin fluctuations 
are broader in momentum space but basically commensurate at ${\bf Q_{AF}}$
(see section {\bf 4}). More importantly, INS experiments in YBCO now 
reported for ten years (first observation in 1991 \cite{rossat91}) 
a sharp magnetic {\it resonance} peak at ${\bf Q_{AF}}$ {\it only} 
in the superconducting (SC) state, which likely results from $d$-wave 
symmetry of the superconducting order parameter \cite{tony,revue-cargese}.
Despite many efforts, this remarkable feature is absent in the spin 
spectrum of LSCO. Only recently \cite{bi2212,he}, a third 
high-$T_c$ system $\rm Bi_2 Sr_2 Ca Cu_2 O_{8+\delta}$ (BSCO) has 
been investigated and, up to now, only in the superconducting state. 
Results in BSCO are fully consistent with YBCO for equivalent doping 
level.

In conventional superconductors, electron-phonon interaction 
was most directly evidenced by tunneling experiments which was 
reproducing the phonon density of states (DOS) \cite{schrieffer}. 
In copper oxides, a similar evidence of the mechanism 
responsible for the high-$T_c$ superconductivity is still missing. 
However, it has been recently argued \cite{campuzano,chubukov} that 
single particle spectrum (at finite wavevector measured by Angle 
Resolved Photoemission (ARPES) or integrated in momentum 
space in tunneling experiments) exhibits features in the superconducting 
state which map the resonance peak in the spin excitation spectrum. 
Independently, a similar connection has been inferred from infrared 
measurements~\cite{carbotte,munzar} showing that conducting carriers 
are strongly coupled to this magnetic mode. This is actually
putting the resonance peak seen in INS experiments
as a clearcut manifestation of a superconductivity mechanism based 
on magnetic fluctuations for the high-$T_c$ in these materials.

In the present lecture, the INS experimental situation will not be
presented in details: this has been already done in LSCO \cite{kastner} 
and in YBCO \cite{revue-cargese,revue-lpr,dai99,tony2000} (a critical 
examination of the INS results in YBCO and BSCO has been recently given 
in Ref. \cite{klosters}). The objective is here rather to show how 
the magnetic INS measurements can give insight about the role of AF 
fluctuations for the high-$T_c$ mechanism. The energy dependence of the 
spin susceptibility will be discussed in both the normal and the 
SC states in section {\bf 3}, and the momentum dependence in 
section {\bf 4}. Before tackling these aspects, let first 
emphasize the anomalous phase diagram empirically inferred from many 
different measurements.

\begin{figure}[t]
 \epsfig{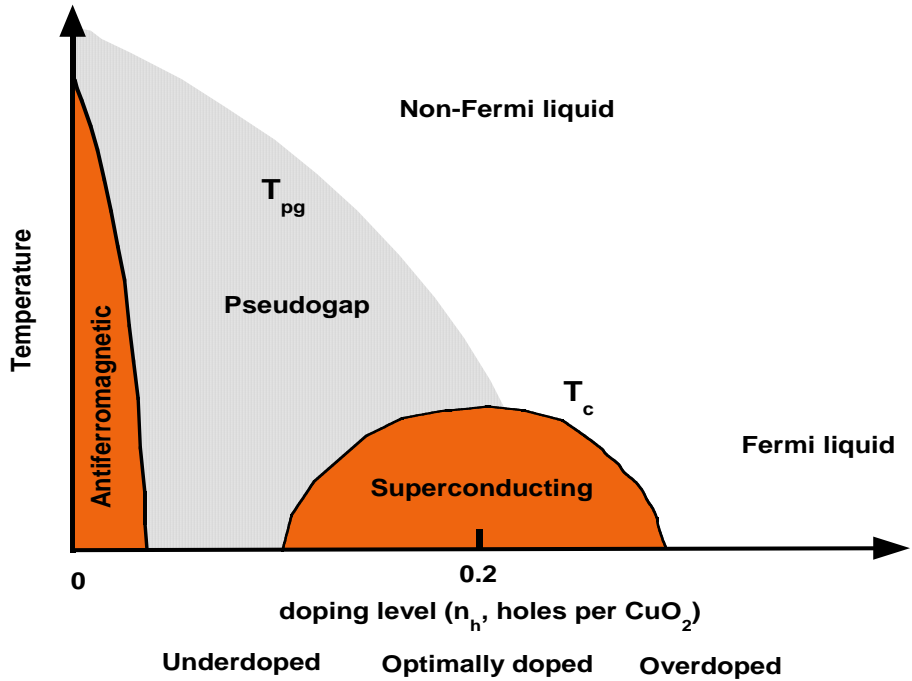}
\caption[xxx]{Generic phase diagram of High-$T_c$ cuprates (from
\cite{batloggvarma}). \label{fig:phasediagram}}
\end{figure}

\section{The High-$T_c$ Cuprates phase diagram}

The high-$T_c$ cuprates are layered systems, systematically  containing 
one or more CuO$_2$ planes, known to be responsible for the high-$T_c$ 
superconductivity. The physical properties of all high-$T_c$ 
Cuprates are strikingly controlled by the number of doped holes, $n_h$,
within the CuO$_2$ planes. Part of the difficulty to describe 
the physical properties is related to that bidimentionnal 
nature of cuprates. A generic phase diagram is shown in Fig. 
\ref{fig:phasediagram} versus doping level \cite{batloggvarma,orenstein}. 
The undoped system is an insulator with a long range ordered N\'eel state
whose propagation wavevector is ${\bf Q_{AF}}$. 
This AFM state disappears with small amount of doped holes, $n_h \sim$ 
2-3 \%, and superconductivity occurs upon increasing further $n_h$. 
The SC transition temperature is going through a maximum for some doping,
{\it so-called} optimal doping, separating an underdoped regime from an 
overdoped regime which exhibit quite different physical properties.

Superconductivity in cuprates is unconventional with a $k$-dependent 
SC order parameter of d$_{x^2-y^2}$-wave symmetry, 
$\Delta_{\bf k}=\Delta_{max}/2 \ (\cos k_x - \cos k_y)$ (where $k_x$ and $k_y$ 
are in-plane wavevectors, $\Delta_{max}$ is the maximum of the SC $d$-wave gap). This has been  very well documented by crossing 
ARPES experiments \cite{shen1} and Josephson effect 
experiments~\cite{tsuei}. A ratio $2 \Delta_{max}/k T_c$ is found $\sim$ 
7$-$9 at optimal doping (see also electronic Raman scattering \cite{hackl})
far from the expected BCS value of $\sim$ 3.5. Further, the single particle 
spectrum shows a doping dependent gap which increases for lower doping 
\cite{mesot}. However, other measurements of superconducting properties 
(Andreev reflection, penetration depth,...) does not follow that trend 
of the single particle gap \cite{deutscher} and rather indicate a second 
lower gap, $\Delta_c$, which actually follows $T_c$ as 
$2 \Delta_c/k T_c \sim$ 5$-$6 for any doping. This second feature has 
been interpreted as a coherence gap \cite{deutscher} but, alternatively, 
it can also arises from different portions of the Fermi surface than the 
single particle gap.

In principle, an insulating-metal transition should occur at low doping before 
superconductivity sets up. However, unusual transport properties for a metal 
are observed in most of the phase diagram indicating a non-Fermi liquid 
behavior, for instance characterized by a linear thermal dependence of the 
resistivity \cite{ito,moschalkov}. This has been subsequently confirmed by 
the absence above $T_c$ of well-defined quasiparticles in ARPES 
spectroscopy \cite{norman}, questioning about the existence of a Fermi 
surface. The strangest part of this phase diagram is certainly the 
{\it pseudogap} state. Indeed, transport measurements \cite{ito,moschalkov}
as well as thermodynamics \cite{loram97} display anomalies (depression) 
below some temperature, referred as $T_{pg}$ in Fig. \ref{fig:phasediagram}, 
much higher than $T_c$. Further, 
various charge spectroscopies - ARPES experiments \cite{ding}, optical 
conductivity \cite{basov}, Raman scattering~\cite{hackl,raman} or tunelling 
experiments \cite{effettunnel} - have reported below $T_{pg}$ (but still 
above $T_c$) a loss of low energies electronic states suggesting 
an opening of a pseudogap. Magnetic properties also affected below 
$T_{pg}$. It is indeed known since 1989 that the uniform spin 
susceptibility as measured by NMR Knight shift does not behave as
a Pauli susceptibility but systematically exhibits a pronounced 
decrease down to lower temperature \cite{alloul}. Copper NMR \cite{nmr}
as well as INS experiments \cite{rossat90} have also evidenced a 
depression of low energy fluctuations at $(\pi,\pi)$ already in the 
normal state, characteristic of a spin pseudogap behavior.
Finally, it should be noticed that all anomalies in the physical 
properties occur, for a fixed doping level, over a certain temperature
range rather than at a well-defined temperature. This is either related 
to the fact that 
each physical property might be differently sensitive to the growing 
of the pseudo gap, or that the chosen definition of the anomalous 
temperature for each experimental technique is quite ambiguous.

This pseudo-gap phase shows the failure of conventional 
Fermi liquid to describe the high-$T_c$ cuprates. Further, the debate 
has been enriched by the observation of static spin and charge ordering 
in non superconducting cuprates ${\rm La_{2-x-y}Nd_ySr_xCuO_{4}}$ 
\cite{tranquada}. This has interpreted as an evidence of inhomogeneous
distribution of charge carriers within the CuO$_2$ plane: the charge
carriers would form lines, yielding the so-called {\it stripes phase},
separating antiferromagnetic regions free of charges.
It has been then speculated that dynamical stripes occur in SC 
cuprates \cite{stripes}, so that the cuprates ground-state is a 
doped insulator without formation of a Fermi surface.
In such a case, the pseudo-gap phase is understood as a 
formation of preformed superconducting pairs without long range 
coherence \cite{orenstein,stripes}. In approaches which still consider
the Fermi surface, the pseudo-gap line is described either as simply 
a crossover line of physical properties \cite{pines,rice},  
or as the trace of a quantum critical point (QCP) \cite{OP1,varma,CLMN} at 
optimal doping, with even a broken symmetry of an hypothetic hidden order 
parameter~\cite{varma,CLMN}. Actually, the scaling of several physical
properties at a common critical doping suggests the occurrence of a 
QCP in the lightly overdoped regime \cite{tallon}. 
The doping dependence of the pseudo-gap, 
sketched as $T_{pg}$ in Fig.~\ref{fig:phasediagram}, actually corresponds 
to that of the single particle gap measured in the SC state \cite{mesot}.
The description of the physics behind the pseudo gap phase then 
would necessarily be important for the understanding of the high-$T_c$
superconductivity. In any case, the unusual phase diagram of
Fig. \ref{fig:phasediagram} suggests an unconventional SC pairing.

\section{Spin susceptibility}

All INS experiments in the high-$T_c$ cuprates have unambiguously 
established the existence of strong antiferromagnetic correlations 
in both the normal and the superconducting states. This drastically 
differs from what we would expected in any conventional superconductor
in both the normal 
state and the superconducting states. Typically, the spin susceptibility 
in standard paramagnetic metals (like Al, Cu,...) would be very weakly 
peaked in momentum space and extends up to high energy (limited by the 
electronic bandwidth, $t \sim$~0.5 eV), as a result of weak electronic 
correlations. The electronic spin fluctuations are then too weak to be 
detected in INS experiments. The amplitude of the magnetic fluctuations 
of several 100 $\mu_B^2$/eV 
observed in the metallic state of cuprates (at the AF wave vector, 
$(\pi,\pi)$, in YBCO \cite{revue-cargese,tony2000} or near $(\pi,\pi)$ 
in LSCO~\cite{aeppli}) is actually much 
larger than what expected in non-interacting metals of the order of 
$1/t \sim 2\mu_B^2$/eV. This, by itself, suggests the importance of 
antiferromagnetism in the microscopic description of the high-$T_c$ 
superconductors. The complete energy, wave vector and  temperature 
dependences of the imaginary part of the spin susceptibility, 
$Im \chi ({\bf Q},\hbar\omega)$, have been reported in YBCO for several dopings
\cite{revue-cargese,revue-lpr,tony2000}. Let us first describe its
behavior in the normal state.

\begin{figure}[t]
\epsfig{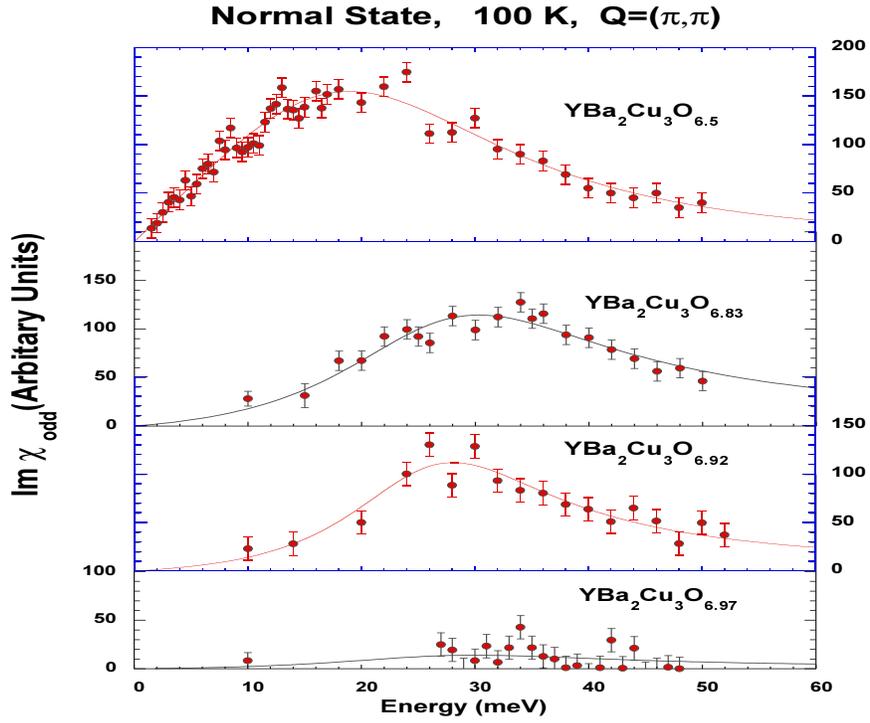}
\caption[suscep_ns]{Normalized imaginary part of the spin susceptibility 
at the AF wavevector in the normal state, at T= 100 K, for four oxygen 
contents in YBCO ($T_c$=45,85,91,92.5 K for $x$=0.5,0.83,0.92,0.97 
respectively). These curves have been normalized to the same units 
using standard phonon calibration \cite{tony2000} (100 counts in the 
vertical scale roughly correspond to $\sim$ 350 $\mu_B^2$/eV i 
absolute units) (from \cite{revue-cargese}). \label{fig:imchi} }
\end{figure} 

\subsection{Normal state}

In a superconducting mechanism based on magnetism \cite{pines,afmodel}, 
the effective interaction would be directly proportional to the spin 
susceptibility measured {\it in the normal state}. The determination 
of the neutron spectrum above $T_c$ is then of great importance. 
The spin susceptibility at T=100 K is reported in Fig.~\ref{fig:imchi} 
at the AF wavevector for 4 different oxygen contents in YBCO corresponding 
to 4 doping levels in Fig. \ref{fig:phasediagram}: $x$=0.92 is 
assumed to be slightly below the optimal doping, $x$=0.5 and $x$=0.83 
are on the underdoped side and $x$=0.97 is slightly on the overdoped 
side (see refs. \cite{revue-cargese,revue-lpr} for details).
Apart for $x$=0.97, $Im \chi ({\bf Q_{AF}},\hbar\omega)$ systematically 
exhibits a maximum around a characteristic energy, $\sim$ 25-30 meV.
Actually, such energy dependences remind that of paramagnons in metallic 
systems where strong electronic interactions enhance, at low energy, the 
bare spin susceptibility \cite{white}.

\begin{figure}[t]
\epsfig{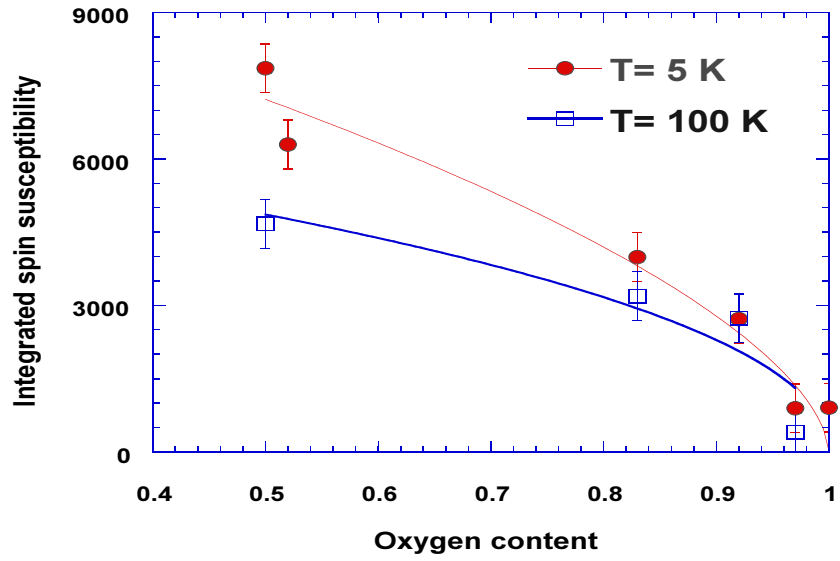}
\caption[sumup]{Doping dependence of 
$\int_0^{50 meV} Im \chi ({\bf Q_{AF}},\hbar\omega) d \omega$ in both the normal 
state (T=100 K from the energy dependences of Fig. \ref{fig:imchi})
as well as in the SC state (the energy dependences of 
$Im \chi ({\bf Q_{AF}},\hbar\omega)$ at T=5 K are from \cite{revue-cargese}). 
Lines are only guides to the eye.
\label{fig:sum} }
\end{figure} 

Interestingly, a drastic decreasing of the AF susceptibility amplitude
is found as a function of doping. This effect is particularly pronounced
near optimal doping where $Im \chi ({\bf Q_{AF}},\hbar\omega)$ is at least 
4 times weaker for $x$=0.97 than $x$=0.92 corresponding to a small 
change of the doping level and for very similar $T_c$. This result
is quite surprising as compared with the copper spin-lattice relaxation 
rate of NMR measurements \cite{nmr}, $^{63}T_1$, which remains
roughly constant from YBCO$_{6.92}$ to YBCO$_7$ at T=100 K 
as, in principle, both probes measure spin fluctuations peaked around
$(\pi,\pi)$. This might indicate either that NMR is sensitive to another
source of magnetism \cite{varmanmr} too broad to be detectable in 
INS experiments or that the spin susceptibility 
has different momentum or energy dependences in the overdoped regime.
[For instance, if the maximum of $Im \chi ({\bf Q_{AF}},\hbar\omega)$ is 
shifted down to $\sim$ 20 meV, magnetic scattering can be 
partially occult by the large nuclear background in the INS spectra.]
At present, it is premature to resolve this alternative:
more accurate INS measurements in YBCO$_7$ are necessary to remove 
that difficulty. 

However, to underline the observed doping dependence of Fig.
\ref{fig:imchi}, a partial energy integration of the spin susceptibility 
at $(\pi,\pi)$, $\int_0^{50 meV} Im \chi ({\bf Q_{AF}},\hbar\omega) d \omega$, 
is shown in Fig. \ref{fig:sum} as a function of the oxygen content.
It should be stressed that Fig.\ref{fig:sum} does {\it not} represent
any sum-rule, but just display the doping dependence of the spectral weight 
of the measured AF dynamical correlations within an arbitrary energy range,
which is nevertheless the most interesting spectral region.
At first glance, Fig. \ref{fig:sum} suggests that AF fluctuations 
cannot be important for the high-$T_c$ mechanism as the magnetic 
fluctuations seem to vanish for samples where $T_c$ is almost as 
large as the maximum $T_c$. But actually, assuming the same momentum 
dependence for YBCO$_{6.97}$ as that for YBCO$_{6.92}$, the upper limit
of the spin susceptibility reported in Fig. \ref{fig:imchi} for 
YBCO$_{6.97}$ is $\sim$ 80 $\mu_B^2 / eV$ at $(\pi,\pi)$ i.e. 
still $\sim$ 20 times larger than the uniform susceptibility 
measured by macroscopic susceptibility or deduced from NMR knight 
shift\cite{mmp}. Therefore, Fig. \ref{fig:sum} does not contradict 
the proposal that electronic interactions are responsible for the 
high-$T_c$ superconductivity \cite{pines,afmodel}. AF fluctuations 
can be still large enough to give rise to a sizeable $T_c$.
Further, it is instructive to mention that the doping dependence of 
AF fluctuations spectral weight reminds that of $T_{pg}$ in Fig. 
\ref{fig:phasediagram}. The existence of strong AF fluctuations 
seems then directly related to the opening of the pseudo-gap, as for
instance it has been suggested within QCP scenario \cite{CLMN,tallon}. 

\subsection{Superconducting state}

The spin dynamics in cuprates exhibit a sharp resonance peak in the SC 
state at some well-defined energy
\cite{rossat91,tony,revue-cargese,revue-lpr,mook,dai99,science,tony2000,bi2212,he} 
which {\it disappears} in the normal state for all doping \cite{klosters}. 
So far, the resonance feature has been found for only two systems, YBCO 
and BSCO, and not in LSCO. As the two former families contain bilayers 
of CuO$_2$ planes in each unit cell, it is not fully established that 
this phenomenon can be generalized to all cuprates. However, the absence 
of the resonance peak in the single layer LSCO system can be caused by 
few reasons, (i) too small $T_c$ ($T_c^{max}=$ 38 K in LSCO) (ii) too much 
disorder \cite{revue-cargese,klosters}, so, its non-observation in LSCO 
does not necessarily mean that the  resonance peak phenomenon will not 
exist in other single CuO$_2$ layer cuprates with higher 
superconducting temperature.

In both underdoped and overdoped regimes, the 
resonance energy, $E_r$, is basically proportional to $T_c$ as 
$E_r/k_BT_c \sim$ 5$-$5.5 \cite{revue-cargese,tony2000,he}. Therefore,
the resonance energy surprisingly does not follow the doping dependence of 
the single particle gap, but actually rather matches the doping dependence 
of the second gap behavior $\Delta_c$ \cite{deutscher}. This is 
certainly an important issue which needs further investigations.
Similarly to the normal state, the overall spectral weight in the SC state 
decreases with increasing doping (Fig. \ref{fig:sum}). However, the 
absolute spectral weight related to the resonance peak itself remains 
approximately constant over the same doping range \cite{dai99,tony2000}.
The resonance peak is then a better defined feature for samples with 
high $T_c$ where the normal state peak becomes weaker.

The occurrence of the resonance peak has motivated a large 
theoretical activity (see e.g. \cite{collective,OP2} and other references in 
\cite{revue-cargese,revue-lpr,tony2000,science,klosters,bi2212,he}).
The simplest approach is to consider the spin response in a BCS 
superconductor within a Fermi-liquid approach \cite{schrieffer}.
In such a case, the resonance peak {\it primarily} results from 
electron-hole pair production across the SC energy gap. 
[In that sense, the resonance peak occurs due to coherence effects in 
a $d$-wave superconductor in a similar way as, 
in conventional superconductors with isotropic $s$-wave gap, the 
Hebel-Slichter peak is observed in NMR experiments \cite{schrieffer}.]
In such a case, the resonance peak is described as an excitation which
creates two quasi-particles at the Fermi level whose momentums 
differ by exactly $(\pi,\pi)$. This requires a d$_{x^2-y^2}$-wave 
symmetry of the SC gap function as $\Delta_{\bf k}$ should change sign from 
any ${\bf k}$ wavevectors to ${\bf k+Q_{AF}}$ \cite{tony}.
The spectrum of this excitation exhibits a gap, corresponding to the 
threshold of the electron-hole  continuum, $\omega_c$, defined by the
minimum of $\sum_{\bf k} (E_{\bf k} + E_{{\bf k+Q_{AF}}})$ (where 
$E_{\bf k}=\sqrt{\epsilon_{\bf k}^2+\Delta_{\bf k}^2}$ 
is the quasi-particle energy in the SC state, $\epsilon_k$ is the 
electronic dispersion in the normal state). This threshold can actually 
be experimentally determined from ARPES measurements. For instance, 
let us consider the case of BSCO at optimal doping: the SC gap is measured 
to be $\Delta_{max}$= 35 meV \cite{mesot}. A close inspection of the 
Fermi surface \cite{fermisurface} shows that the quasi-particles which 
are connected by the $(\pi,\pi)$ momentum have the energy 
$E_k  \simeq  0.9 \Delta_{max}$ at the Fermi level. The electron-hole 
continuum is then, $\omega_c \simeq 1.8 \Delta_{max}  \simeq  63$ meV. 
The resonance energy for that composition has been reported  at 
$E_r$= 43 meV \cite{bi2212}, clearly lower than $\omega_c$. The above 
scenario of non-interacting spin susceptibility, $\chi_0$, is then not 
enough to account for the observed sharp peak. A ratio 
$E_r \simeq 1.2 \Delta_{max}$ is found indicating that the resonance 
peak occurs well below the electron-hole continuum. Within a 
Fermi-liquid-like approach, this experimentally shows that the resonance 
peak is a true collective mode of $d$-wave superconductivity, corresponding
to the strong coupling limit \cite{chubukov}. Interactions, like $J({\bf Q})$, 
in a RPA scheme actually produces this collective excitation under the 
condition that, $1 - J({\bf Q_{AF}}) Re \chi_0$= 0 \cite{collective,OP2}. 
A momentum dispersion of this collective mode has been also theoretically 
predicted \cite{OP2} in agreement with recent INS reports 
\cite{science,klosters}.

\subsection{Spin dynamics and single particle excitation}

As mentioned in the introduction, the unusual spectral lineshape, known as
the {\it peak-dip-hump} structure \cite{campuzano}, of the quasi-particles 
measured by ARPES in the superconducting state of BSCO has been interpreted
as a result of a coupling with collective excitations centered at the AF 
momentum \cite{shen2} and more specifically with the magnetic resonance 
peak \cite{norman2}. Namely, quantitative correspondence of the resonance 
energy with the energy separated the peak and the dip have been proposed 
through electron-electron contribution to the electronic self energy 
\cite{campuzano,chubukov}. This proposal, discussed in details in this 
book by J. Mesot, seems to properly agree with the neutron 
data \cite{campuzano}.

In many aspects, this proposal is similar to the recent claim of 
superconductivity mediated by spin fluctuations in the heavy-fermion 
compound, UPd$_2$Al$_3$ \cite{huth}. This system is remarkable as it 
simultaneously exhibits an AF N\'eel state below $T_N=14.3 K$ and a 
superconducting state below $T_c \simeq 2 K$ with also an unconventional
symmetry of the SC gap. Tunneling spectroscopy in the SC state \cite{huth} 
displays, above the SC gap, oscillations at energies comparable 
to the  spin-wave energy $\sim$ 1.4 meV, which has been directly 
measured by INS \cite{metoki,bernhoeft} at the propagation wavevector
 of the AF structure, $Q_0=(0,0,{1\over2})$.
The similarity with tunneling data showing the phonon DOS in conventional 
superconductors is striking and suggests a magnetic origin of the
superconductivity in that system. Further, a new magnetic excitation 
appears in the neutron spectrum {\it only} in the SC state at an energy,
$\hbar\omega \simeq$ 0.36 meV, and at 
$Q_0$~\cite{metoki,bernhoeft}. This peak has also other common features 
with the resonance peak in the cuprates as it occurs at an energy lower 
than twice the SC gap at the wavevector characteristic of the magnetic 
correlations. Therefore, the similarity with cuprates is quite 
striking, and the main difference with the proposal made in the high-$T_c$ 
cuprates is that, in UPd$_2$Al$_3$, the tunneling spectrum shows the 
pre-existing normal state spin excitation spectrum (not the additional 
"resonance" peak of the SC state).

The next question is then: in the cuprates, what might be the link between 
the resonance peak below $T_c$ and the normal state maximum of the 
spin susceptibility ? For instance, it has 
been proposed that the normal state peak observed in neutron 
scattering is a precursor peak of the resonance peak~\cite{morr}. 
However, the normal state maximum has not the same doping dependence 
as the resonance peak \cite{revue-cargese}. Especially, spin 
susceptibilities in underdoped YBCO$_{6.83}$ and in optimally doped 
YBCO$_{6.92}$ exhibit a maximum at $\hbar\omega \sim$ 
30 meV in the normal state (see Fig. \ref{fig:imchi}) whereas they are 
characterized by two distinct resonance energies of 35 and 41 meV, 
respectively \cite{revue-cargese}. Independent physical process are 
likely needed to account for the two characteristic energies 
(a possible scenario is given in \cite{manske}).
Nevertheless, both features are also necessarily inter-related as, 
for instance, the resonance intensity is formed from the broad 
normal state feature. Further, they systematically occur in a very 
similar energy range which makes a clear difference
with the heavy-fermion compound, UPd$_2$Al$_3$, where in the SC state
the two  excitations - spin-waves and the additional "resonance'' peak - 
occur in two distinct energies ranges. Therefore, the fact that,
in the cuprates, the resonance peak in the SC state replaces 
the broad normal state peak - affecting both 
the energy \cite{revue-cargese,revue-lpr} and the momentum 
\cite{science,klosters} lineshapes of the spin susceptibility - might 
explain why the ARPES anomalous lineshape is sensitive to the resonance 
energy. Actually, the energy between the peak and the dip is also 
within errors fully consistent with the normal state maximum of the spin 
susceptibility. Further, to complete the comparison, it should be stressed 
that the magnetic fluctuations in UPd$_2$Al$_3$ are typically thought 
to arise from two different subsystems characterized, aboved $T_c$ by 
two coupled modes \cite{bernhoeft} whereas, in cuprates, a single 
broad response is observed (see Fig. \ref{fig:imchi}).
 
\section{Momentum dependence}

The momentum dependence the peak in the spin susceptibility in the 
normal state exhibits also a very interesting feature. Indeed, in 
the underdoped regime of both LSCO \cite{yamada} and YBCO \cite{balatsky}, 
the superconducting $T_c$ is {\em linearly} proportional to 
the typical momentum extension of the AF fluctuations.

In the case of LSCO, the low energy spin fluctuations are peaked 
at some wavevectors displaced by a doping dependent amount, $\delta$. 
 $\delta$ is temperature independent and does not depend on 
energy, a least up to $\sim$ 20-25 meV (the question whether the 
fluctuations remain incommensurate up to higher energy or becomes 
commensurate is actually controversial). This discommensuration is 
then characteristic of {\it only} the doping level. However, both 
features are not simply proportional each other as $\delta$ saturates at 
high doping, rather following the doping behavior of $T_c$ \cite{yamada}.

\begin{figure}
 \epsfig{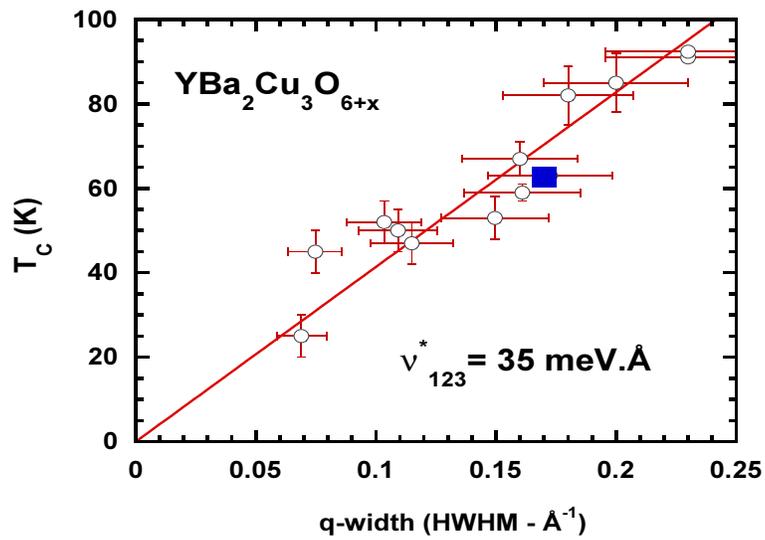}
\caption[xxx]{ Superconducting transition versus the half width at half 
maximum  of the peak in the spin susceptibility, $\Delta_q$
    (from \cite{balatsky}). \label{fig:TcvsDq}}
\end{figure}

In YBCO, the spin susceptibility in the normal state is basically
commensurate. [Incommensurability in YBCO have been also reported 
\cite{mook} but they predominantly occur in the SC state, and actually, 
belong to the same excitation as the resonance peak 
\cite{science,klosters}. The different temperature, energy and doping 
dependences of the discommensurations in both YBCO and LSCO basically 
call for a different interpretation.] Interestingly, the typical momentum
extension 
of commensurate AF fluctuations in YBCO, the half width at half maximum  
of the peak in the spin susceptibility, $\Delta_q$, behaves very similarly 
to the discommensuration parameter in LSCO \cite{balatsky}. $\Delta_q$ is 
also found temperature independent
\cite{revue-lpr} and very weakly energy dependent \cite{tony2000}.
As $\delta$ in LSCO, $\Delta_q$ is then related to the doping level
in YBCO and a linear relationship, $T_c = \hbar v^* \Delta_q$ for a large 
number of INS experiments is also found (Fig. \ref{fig:TcvsDq}) 
\cite{balatsky}, 
similar to the linear relation between incommensurate peak splitting 
in LSCO.

The exact meaning of this relation remains unclear and, so far, has been 
discussed within the stripes picture as an evidence of charged stripes
formation in all cuprates \cite{balatsky,orenstein,stripes}. Other 
interpretations have to be considered as, on general grounds, this linear 
relation indicates that $T_c$ in the underdoped cuprates is controlled by 
the momentum extension of the AF fluctuations in the normal state. 
Superconductivity is indeed limited
by the strength of the magnetic correlations, their amplitude at 
$(\pi,\pi)$ as shown by Fig. \ref{fig:sum}, but also by their location 
in momentum space: broader AF correlations, better $T_c$. Actually, a 
relation between $T_c$ and the inverse magnetic correlation length, 
$\xi^{-1}$ (which basically corresponds to $\Delta_q$),
has been recently discussed in a spin-fermion model where  
superconductivity is magnetically induced \cite{chubukov2}.
More works are certainly needed in that direction.

\section{Conclusion}

Finally, the occurrence of strong and doping dependent antiferromagnetic 
fluctuations as well as their close link with the unusual physical 
properties of high-$T_c$ superconductors and with the SC 
temperature itself \cite{revue-cargese,he,balatsky} naturally 
militates in favor of an electron-electron origin for the SC pairing.
Further, the magnetic resonance peak is proposed to be responsible 
for the anomalous shape of the microscopic electronic properties 
(single particle spectrum and optical conductivity). If this
idea is correct, it gives a clearcut signature of the high-$T_c$ 
pairing based on antiferromagnetism (by analogy with the tunneling 
experiments in conventional superconductors which were reproducing 
the phonon DOS).

\section*{Acknowledgments}

I am grateful to my collaborators, Bernard Hennion and Yvan Sidis
(Laboratoire L\'eon Brillouin, Saclay), Bernhard Keimer (Max Planck 
Institute, Stuttgart), Louis-Pierre Regnault (CEA Grenoble, Grenoble), 
Sascha Ivanov (Institut Laue Langevin, Grenoble) and Jacques Bossy
(CRTBT-CNRS, Grenoble). I also would like to acknowledge stimulating 
discussions  with Sascha Balatsky, Nick Bernhoeft, Joel Mesot, 
Flora Onufrieva, Pierre Pfeuty, and Jeff Tallon.

\end{document}